\documentclass[10pt,conference]{IEEEtran}

\usepackage{cite}
\usepackage{amsmath,amssymb,amsfonts}
\usepackage{graphicx}
\usepackage{booktabs}
\newcommand{\rowsep}{\midrule[0.3pt]} % thin rule between table rows
\usepackage{tabularx}
\usepackage{multirow}
\usepackage{makecell}
\usepackage{enumitem}
\usepackage{adjustbox}
\usepackage{xcolor}
\usepackage{xspace}
\usepackage{url}
\usepackage[hidelinks]{hyperref}
\usepackage{ifthen}
\usepackage{pifont}
\usepackage[strict]{changepage}
\usepackage{framed}
\usepackage{tikz}
\usetikzlibrary{calc,positioning,arrows.meta,fit}

\newcommand{\gb}{IoTPwnBench\xspace}
\newcommand{\NDevices}{11\xspace}
\newcommand{\NImages}{716\xspace}
\newcommand{\NTwins}{716\xspace}

\newcommand{\NTotalImages}{1432\xspace}
\newcommand{\NScen}{86\xspace}
\newcommand{\NCWE}{28\xspace}
\newcommand{\NShared}{63\xspace}
\newcommand{\NSpecific}{23\xspace}
\newcommand{\NModules}{56\xspace}
\newcommand{\NOptIn}{517\xspace}  % images that need an explicit opt-in flag to build
\newcommand{\NPerDev}{64\xspace}   % scenarios on each non-router device (63 shared + 1 specific)
\newcommand{\NRouter}{76\xspace}   % scenarios on the router (63 shared + 13 specific)
\newcommand{\Arch}{\texttt{qemuarm64}\xspace}

\newboolean{showcomments}
\setboolean{showcomments}{true}
\ifthenelse{\boolean{showcomments}}
  {\newcommand{\mynote}[2]{\textcolor{gray}{\footnotesize\textbf{[#1: #2]}}}}
  {\newcommand{\mynote}[2]{}}
\definecolor{promptbg}{HTML}{E6F3FF}
\definecolor{promptborder}{HTML}{2166A5}
\definecolor{answerbg}{RGB}{225,255,225}
\definecolor{answerborder}{RGB}{0,140,75}

\begin{document}

\title{IoTPwnBench: A Synthetic Benchmark for IoT Firmware Security Analysis}

%%%%% Authors %%%%%

\author{
\IEEEauthorblockN{Luca Ferrari}
\IEEEauthorblockA{
\textit{IMT School for Advanced Studies Lucca}\\
Lucca, Italy\\
luca.ferrari@imtlucca.it
}

\and
\IEEEauthorblockN{Marco Alecci}
\IEEEauthorblockA{
\textit{SnT, University of Luxembourg}\\
Luxembourg, Luxembourg\\
marco.alecci@uni.lu
}

\and
\IEEEauthorblockN{Luca Verderame}
\IEEEauthorblockA{
\textit{University of Genoa}\\
Genova, Italy\\
luca.verderame@unige.it
}

% New row: center the remaining two authors
\and[\hfill\mbox{}\par\vspace{0.8em}\mbox{}\hfill]

\IEEEauthorblockN{Jacques Klein}
\IEEEauthorblockA{
\textit{SnT, University of Luxembourg}\\
Luxembourg, Luxembourg\\
jacques.klein@uni.lu
}

\and
\IEEEauthorblockN{Francesco Pagano}
\IEEEauthorblockA{
\textit{DistriNet, KU Leuven}\\
Leuven, Belgium\\
francesco.pagano@kuleuven.be
}
}

%%%%% Abstract and keywords %%%%%
\maketitle

\begin{abstract}
  Building and evaluating security tools for IoT firmware requires targets whose weaknesses are known in advance and can be reproduced in a running device. Two problems remain: (1) firmware collected from real-world products provides realistic software, but hardware dependencies and undocumented defects make it difficult to establish what a security tool should be expected to find; and (2) existing benchmarks typically collect sets of vulnerable or misconfigured firmware images without providing purpose-built vulnerable services with explicit weaknesses that a security analysis tool can target.

We present \emph{\gb}, a synthetic, modular benchmark that brings these requirements together. 
\emph{\gb} contains \NImages vulnerable firmware images covering \NScen  weakness scenarios and \NCWE CWE classes across 11 device profiles, including industrial controllers, health monitors, and household devices. 
%The device profiles include industrial controllers and health monitors alongside household devices. 
Each vulnerable image runs a plausible device application with simulated sensors and actuators that generate live readings over its native interfaces while exposing one target weakness mapped to a primary CWE. 

A documented exploitation payload and observable signal establish the expected behavior independently of the evaluated tool. Every vulnerable image is paired with a corrected twin that preserves the device and service while removing that weakness. Researchers can therefore compare the same target before and after correction to assess detection and false positives for the target weakness. The benchmark produces kernel and filesystem binaries for static inspection and dynamic testing under QEMU. Device profiles and weakness modules remain separate, allowing the same scenario to be reused across device contexts. 
%The catalog contains \NImages vulnerable images and \NTwins corrected twins across \NDevices device classes, covering \NScen scenarios and \NCWE CWE classes. 
Vulnerable targets were validated manually, and the open source build workflow regenerates the firmware bit-for-bit.
\end{abstract}


%%%%% CONTENT %%%%%
\section{Introduction}

IoT firmware analysis tools must handle devices whose software, peripherals, and communication interfaces differ. Evaluating these tools requires executable targets with known weaknesses, enough device behavior to exercise their interfaces, and corrected counterparts to check their findings. Firmware from deployed devices offers realistic software, but its defects are incompletely documented \cite{10.1145/3428334} and its execution often depends on unavailable hardware \cite{10.1145/3423167,10.1145/3433210.3453093}, which makes controlled comparisons difficult.
% \ma{do we have a paper or something for this sentence?}
Existing benchmarks provide configurable vulnerable systems \cite{10.1145/3600160.3600181}, emulated targets with disclosed CVEs \cite{electronics14050856,electronics12143010}, and deliberately insecure training firmware \cite{owasp2026iotgoat}. B4IoT \cite{10.1145/3600160.3600181} characterizes the weaknesses of Linux-based IoT devices and maps them to CWE and OWASP IoT Top Ten categories, but its images are Alpine containers without a device application. Its weaknesses are misconfigured services, packages with unresolved CVEs, and standalone vulnerable executables that protocol interfaces call directly, and each image combines many of them. A tool evaluated on these images therefore analyzes a container image rather than the firmware of a running device, and a result is hard to trace to a single weakness. We focus on combining device-specific runtime behavior, bootable firmware binaries, and paired vulnerable and corrected variants in a modular catalog.

We present \gb, a synthetic benchmark of \NImages vulnerable firmware images covering \NScen weakness scenarios and \NCWE CWE classes across \NDevices device profiles. The profiles span a network router, an IP camera, an industrial PLC, a health monitor, and smart-home equipment. Each runs a management application whose services expose simulated sensors and actuators. Depending on the profile, these produce physical measurements, actuator and operating states, and counters, which a simulator updates over time and the services expose over HTTP, MQTT, or Modbus/TCP. Dynamic tests therefore meet running applications and changing device state.

The build produces Linux kernels and ext4 filesystem images with the compiled applications, scripts, configurations, and services, which researchers can feed to existing firmware-analysis workflows, inspect statically, or boot under QEMU to test the exposed interfaces. Applications and weaknesses are built on open-source components, so no proprietary vendor firmware is needed. Device profiles and weakness modules are selected independently at build time, so a scenario can be reused across device classes.

Every vulnerable image has a corrected twin, \NTwins in total, that keeps the device profile, service interfaces, and behavior on valid input but removes the target weakness. A pair shows whether a tool detects the weakness and whether it still reports it after the correction, which counts as a false positive. Each vulnerable image has a documented exploitation payload and an observable ground-truth signal, validated manually, and the open-source build regenerates the firmware bit-for-bit.

Our contributions are the following.
\begin{itemize}
    \item A modular benchmark of \NDevices IoT device profiles with simulated peripherals and native services, and \NScen scenarios.
    \item A catalog of \NImages vulnerable firmware targets and \NTwins corrected twins, with documented payloads and signals for evaluating the target weaknesses.
    \item A reproducible build and QEMU workflow with kernel and filesystem binaries for static and dynamic analysis.
\end{itemize}

The benchmark, its build workflow, and the documentation of every scenario are publicly available.\footnote{\url{https://github.com/Mobile-IoT-Security-Lab/IoTPwnBench}}
% \ma{Visto che é un DATA paper metteri già qui alla fine dell'intro il lInk al dataset bello in grande. Anche se ce l abbiamo dopo su Availability}
\section{Dataset Overview}

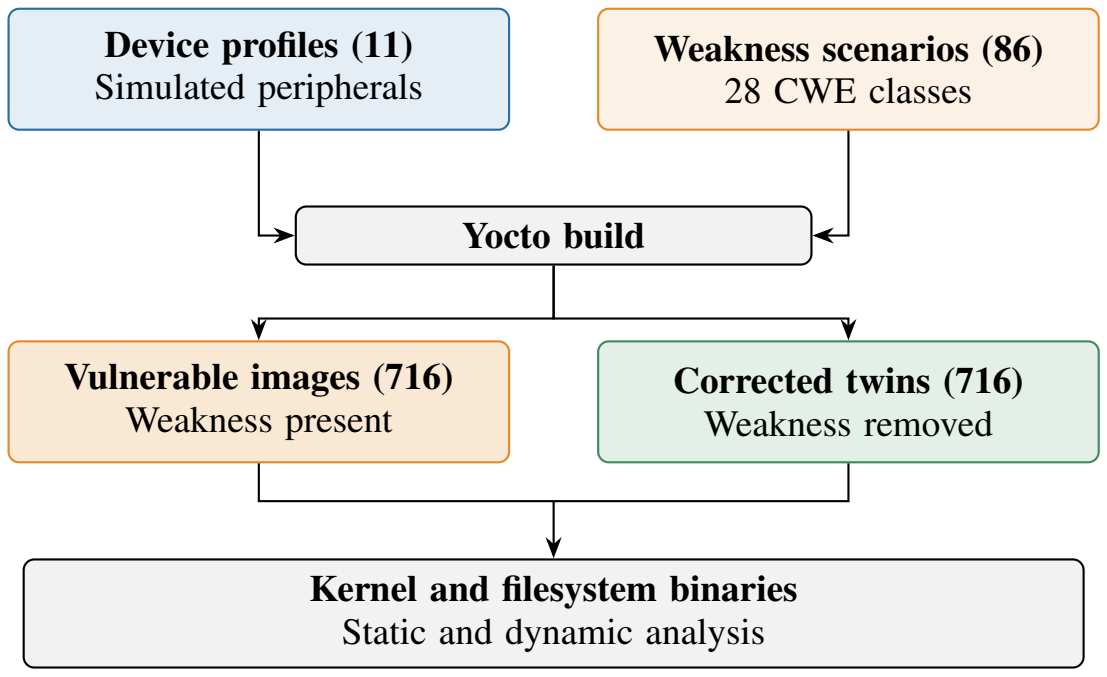
\begin{figure}[ht]
\centering
\begin{tikzpicture}[font=\scriptsize, >={Stealth[length=4pt]},
    blk/.style={draw, rounded corners=2pt, align=center, inner sep=3pt}]
  \definecolor{devcol}{RGB}{38,110,170}
  \definecolor{wkcol}{RGB}{226,138,40}
  \definecolor{gtcol}{RGB}{60,140,90}
  \node[blk, text width=3.1cm, minimum height=0.8cm,
        draw=devcol, fill=devcol!12] (dev) at (0,0)
    {\textbf{Device profiles (\NDevices)}\\ Simulated peripherals};
  \node[blk, text width=3.1cm, minimum height=0.8cm,
        draw=wkcol, fill=wkcol!12] (wk) at (3.9,0)
    {\textbf{Weakness scenarios (\NScen)}\\ \NCWE CWE classes};
  \node[blk, text width=3.2cm, fill=black!5] (build) at (1.95,-1.1)
    {\textbf{Yocto build}};
  \draw[->] (dev.south) |- (build.west);
  \draw[->] (wk.south) |- (build.east);
  \node[blk, text width=3.1cm, minimum height=0.8cm,
        draw=wkcol, fill=wkcol!20] (vuln) at (0,-2.2)
    {\textbf{Vulnerable images (\NImages)}\\ Weakness present};
  \node[blk, text width=3.1cm, minimum height=0.8cm,
        draw=gtcol, fill=gtcol!14] (fixed) at (3.9,-2.2)
    {\textbf{Corrected twins (\NTwins)}\\ Weakness removed};
  \draw[->] (build.south) -- ++(0,-0.35) -| (vuln.north);
  \draw[->] (build.south) -- ++(0,-0.35) -| (fixed.north);
  \node[blk, text width=6.8cm, fill=black!5] (binaries) at (1.95,-3.6)
    {\textbf{Kernel and filesystem binaries}\\ Static and dynamic analysis};
  \coordinate (merge) at ($(vuln.south)!0.5!(fixed.south) + (0,-0.25)$);
  \draw (vuln.south) |- (merge) -| (fixed.south);
  \draw[->] (merge) -- (binaries.north);
\end{tikzpicture}
\caption{Composition of \gb.
Corrected twins retain the device profile and service while removing the target weakness. }
\label{fig:overview}
\end{figure}

\gb contains \NImages vulnerable image recipes and \NTwins corrected twins, \NTotalImages recipes on the reference architecture, \Arch. Each vulnerable image pairs one of \NDevices device profiles with one of \NScen weakness scenarios (Figure~\ref{fig:overview}). The profiles are a network router, an IP camera, a smart plug, a thermostat, a PLC, a smart bulb, a smart hub, an environmental sensor, a smart speaker, a health monitor, and a door lock. \NShared scenarios are instantiated on every profile and \NSpecific target a single device through its native interface, so each non-router profile carries \NPerDev scenarios and the router carries \NRouter, which gives $\NShared \times \NDevices + \NSpecific = \NImages$ vulnerable images. The build supports five QEMU machines (\texttt{qemuarm64}, \texttt{qemuarm}, \texttt{qemux86-64}, \texttt{qemux86}, and \texttt{qemumips}), and each device--scenario pair is counted once regardless of the machine.

A device profile provides the management application, a peripheral specification, and a native service. The router serves a web dashboard over HTTP, the camera serves JPEG snapshots and an MJPEG stream, the PLC speaks Modbus/TCP, and the remaining eight devices publish telemetry over MQTT. A shared simulator reads each specification and updates peripheral values every few seconds, using mean-reverting processes for continuous signals, a seasonal term with noise for daily cycles, Markov processes for actuators and enumerated states, and monotonic accumulators for counters. Derived signals are computed from other signals of the device so that related values stay consistent, and all values appear in the web panel and on the native protocol.

Each scenario implements one target weakness, labeled with a primary CWE, and is added at build time without modifying the device software. The \NScen scenarios come from \NModules weakness modules, since one module can yield a family of variants that expose the same weakness through different files, input patterns, or interfaces. The \NCWE CWE classes cover memory safety errors (CWE-120, 121, 122, 190, 416, 476, 787), injection and input handling (CWE-20, 77, 78, 79, 88, 89, 502), path and link handling (CWE-22, 61), authentication, credential, and cryptographic weaknesses (CWE-287, 294, 306, 327, 352, 521, 798, 862), and permission, exposure, and availability problems (CWE-200, 250, 276, 400). Scenarios also differ in where the weakness surfaces: a dashboard form, a native network service, a configuration file, or a device command.

\begin{table}[ht]
\centering
\caption{Weakness categories with their CWE classes, scenarios, and vulnerable images.}
\label{tab:categories}
\small
\adjustbox{width=0.8\columnwidth}{%
\begin{tabularx}{\columnwidth}{@{}>{\raggedright\arraybackslash}Xrrr@{}}
\toprule
Category & CWEs & Scenarios & Images \\
\midrule
Memory safety errors & 7 & 9 & 49 \\ \rowsep
Injection and input handling & 7 & 19 & 159 \\ \rowsep
Path and link handling & 2 & 8 & 68 \\ \rowsep
Authentication, credentials, cryptography & 8 & 42 & 392 \\ \rowsep
Permissions, exposure, availability & 4 & 8 & 48 \\
\midrule
Total & \NCWE & \NScen & \NImages \\
\bottomrule
\end{tabularx}%
}
\end{table}

Every vulnerable image ships with the ground truth of its target weakness, namely its primary CWE, a documented exploitation payload, and the signal that confirms the weakness triggered. Six modules also map the target weakness to related CWEs, for example a contributing CWE-20 or a broader parent, without separate images. An image may contain other, non-target weaknesses, which the ground truth does not cover. Table~\ref{tab:gt-examples} shows one image per weakness category, and four reproduce public CVEs. Images are named \path{<device>-<cwe>-<slug>}, and corrected twins append \texttt{-fixed}.

\begin{table}[ht]
\centering
\caption{One vulnerable image per weakness category, each on a different device and interface}
\label{tab:gt-examples}
\scriptsize
\renewcommand{\arraystretch}{1.05}
\setlength{\tabcolsep}{3pt}
\begin{tabularx}{\columnwidth}{@{}>{\raggedright\arraybackslash}p{1.03cm}>{\raggedright\arraybackslash}p{1.45cm}>{\raggedright\arraybackslash}X>{\raggedright\arraybackslash}p{2.07cm}@{}}
\toprule
Category & Image & Payload & Signal \\
\midrule
Memory safety & PLC, CWE-121 \cite{cve2018-20818}
  & Modbus function \texttt{0x41} with a 36-byte chunk for a 32-byte field & Status \texttt{STACK\_OVERFLOW} \\ \rowsep
Injection & Router, CWE-78
  & Host field \texttt{127.0.0.1; id} in the diagnostics form & \texttt{uid=995} \texttt{(routerweb)} in the output \\ \rowsep
Path and link & IP camera, CWE-22 \cite{cve2016-15055}
  & \texttt{recording.cgi?file=} with an encoded \texttt{../../etc/passwd} & Contents of \texttt{/etc/passwd} \\ \rowsep
Authenti\-cation & Door lock, CWE-294 \cite{cve2018-16242}
  & Captured \texttt{action=unlock} frame sent twice over MQTT & Same frame accepted twice \\ \rowsep
Exposure & Health monitor, CWE-200 \cite{cve2018-10599}
  & \texttt{READ 32 52} on the monitor TCP service & Patient identifier and vital data \\
\bottomrule
\end{tabularx}
\end{table}
\section{Data Collection}
% \ma{qui mettere giusto una frase di intro tipo Our goal is bla bla. In thi section we explain/describe ...}
Every vulnerable image is built around one target weakness with a documented, reproducible trigger and effect.
We describe how we selected and implemented the weaknesses, built and validated the images, and kept them safe to distribute.

\textit{Weakness selection.} We chose the weaknesses from the characterization of B4IoT \cite{10.1145/3600160.3600181}, whose classes, mapped to the CWE Top 25 and the OWASP IoT Top Ten, cover weakly configured SSH, FTP, and Telnet services, sudo and SUID misconfigurations, cron jobs, hardcoded credentials, password leaks, an SSH backdoor, and vulnerable C, Java, and Bash code reachable over CoAP, MQTT, REST, and web servers.
We used it to choose which weaknesses to cover, not as an implementation base, and reuse no code or image of B4IoT.
Each scenario is a module written for an emulated device that runs its own application, peripherals, and services, and we added \NSpecific device-specific scenarios that act through one device's native interface.
Ten of them are inspired by public CVEs and reproduce the weakness and attack surface of the report on a synthetic device without copying vendor code.
Every target weakness maps to a primary CWE \cite{mitre2026cwe}.

\textit{Implementation.} A scenario is a weakness module, a recipe that installs code, configuration, or a dashboard module into an image, and a device is a profile with the management application, peripheral specification, and native service. A short image recipe selects one profile and one module (Figure~\ref{fig:recipe}), so adding a scenario needs only a module and a few lines. A corrected module is a second recipe that replaces only the files carrying the weakness, and a script generates each twin recipe by selecting it for the same profile.

\begin{figure}[ht]
\centering
\scriptsize
\begin{verbatim}
require recipes-images/iotbench-image.inc
require recipes-devices/<device>/device-<device>.inc
IOTBENCH_CWE = "CWE-<id>"
IOTBENCH_VULN_INSTALL = "<module>"
\end{verbatim}
\caption{Image recipe structure. \texttt{<device>} names the profile, \texttt{<id>} the CWE, and \texttt{<module>} the weakness module. The twin selects \texttt{<module>-fixed}.}
\label{fig:recipe}
\end{figure}

Images are built with the Yocto Project (release 6.0.3) in a Debian 12 container.
Layers are pinned to fixed revisions and a shared cache holds downloads and build state, so the toolchain does not depend on the host, and we verified that rebuilds from the same sources are bit-identical.

\textit{Ground truth and validation.} For every scenario we documented the entry point, the exploitation payload, and the observable signal that confirms the weakness triggered, and 21 modules also ship a machine-readable ground-truth file, which records exact source lines in four of them.
We then built and booted every image under QEMU, triggered its payload manually, and compared the observed signal with the documented one, fixing the module and repeating the check when they differed.
Such differences can come from build settings, for example default compiler hardening that stops a memory corruption payload before its signal.
Scripted regressions cover part of the catalog without booting a guest.

\textit{Safety and distribution.} Because the images are intentionally vulnerable, \NOptIn of them need an explicit opt-in flag that keeps them out of an ordinary build, and the build script sets only the flag each image needs.
The emulator forwards guest services only to the host loopback interface.
\section{Use Cases}

The benchmark is recent, so the uses below are those its design supports, not ones already reported.
% \ma{ci sta dire che il benchmark é recent e quindi parlare solo di utilizzi futuri. Tuttavia metterei comunque qualche riferimento a tool esistenti tipo per dire. EH guarda ora con il nostro bechmark puoi comparare questi due tool sullo stesso benchmark avoid bias etc..}
% \ma{Piu in generalre provere a mettere almeno qualche cit dalla literature qui, per i possibili utilizzi}

\textit{Evaluating firmware analysis tools.} Prior IoT benchmarks were used to compare five static firmware analyzers \cite{10.1145/3600160.3600181} and seven black-box protocol fuzzers \cite{electronics12143010}.
\gb supports comparisons of this kind in which every tool runs on the same images and is scored against the same ground truth, which avoids the bias of evaluations where each study selects its own targets \cite{10.1145/3243734.3243804}.
Each vulnerable image documents its target weakness, primary CWE, payload, and signal. The kernel and filesystem binaries provide inputs for firmware inspection, and QEMU exposes the running device for dynamic testing. The same binaries can be given to frameworks that re-host firmware and instrument its code for dynamic analysis \cite{10.1145/3696630.3728597}, and the documented signal tells whether the instrumented run reached the weakness. Corrected twins allow the same tool to analyze paired versions with and without the target weakness, as the flawed and non-flawed test cases of Juliet do for source code \cite{testjuliet}. A report of that weakness in the corrected implementation can be scored as a false positive. A static analyzer can be scored on whether it reports the documented weakness and, where the ground truth records a source line, the right location, a fuzzer on whether it reaches the signal through the native interface of the device, and a dynamic testing framework on whether it finds the entry point. The score is computed per image and per CWE, and it never uses the verdict of the evaluated tool as the oracle.

\textit{Isolating what drives detection.} The same weakness is instantiated on up to \NDevices device classes, and the same image is built for five machines. A study can therefore change one factor at a time, the device software, the interface, or the architecture, while the weakness stays fixed. This is hard to do on real firmware, where these factors change together.

\textit{Regression and teaching.} A tool pipeline can run on the catalog as a regression suite whose expected findings are known and stable, because every image is rebuilt identically from source. The documented payloads also make the images usable for hands-on exercises in an isolated lab, a role that IoTGoat fills with a single router image \cite{owasp2026iotgoat}.
\section{Limitations and Future Work}

\textit{Limitations.} The weaknesses are designed rather than collected, so they show how a weakness appears in a plausible device and not how often or in what form it occurs in the field. The ten CVE-inspired scenarios reproduce a report on a synthetic device and not the vendor code.
The \NScen scenarios come from \NModules modules, and many are variants of one family, so the number of images overstates the number of independent weaknesses.
Each vulnerable image has a single target weakness, which gives a clean ground truth but leaves out interactions between weaknesses.
Since we wrote the device software and the ground truth covers only the target weakness, other, non-target defects cannot be excluded.
All images are Linux-based and run under QEMU with simulated peripherals, so microcontroller and RTOS firmware and the behavior of real hardware are out of scope.
Deployment on a physical board requires a matching Yocto machine configuration and board support package.
The machine-readable ground-truth files of the 21 modules that have one do not yet share a schema.
Validation of the vulnerable images was manual and only part of the catalog has a scripted regression.

\textit{Future work.} A follow-up study will evaluate static firmware analyzers, for example EMBA \cite{emba2026}, FACT\_core \cite{factcore2026}, and Karonte \cite{9152796}, and dynamic testing tools, for example IoTFuzzer \cite{chen2018iotfuzzer}, Snipuzz \cite{10.1145/3460120.3484543}, and the companion-app penetration testing of MITHRAS \cite{PAGANO2026112918}.
It will ask which weakness classes and device interfaces each tool family misses and how the answer changes across architectures.
Extensions include scoring that separates reaching, triggering, and detecting a weakness as in Magma, more device classes, scenarios that combine weaknesses, and microcontroller targets.
\section{Related Work}
\label{sec:related-work}

\textit{IoT firmware benchmarks.} B4IoT \cite{10.1145/3600160.3600181} builds Alpine container images for x86 and ARMv7, whereas \gb emulates complete devices with simulated peripherals on five QEMU machines and builds each vulnerable image around one target weakness. IoTBenchSL \cite{electronics14050856} and its dataset \cite{a101e2024iotvulbench} reproduce disclosed CVEs in emulated firmware, with 100 firmware and 100 vulnerability benchmarks aimed at fuzzing. IoTFuzzBench \cite{electronics12143010} gathers 14 firmware images and 30 verified vulnerabilities to compare seven black-box protocol fuzzers, and FirmBench \cite{duong2026firmrebugger} offers 313 bug oracles over real monolithic firmware binaries, scored as reached, triggered, or detected for nine fuzzers. These benchmarks build on real firmware, which is realistic but ties them to the licensing and undocumented defects of real images. BenchIoT \cite{8809492} targets microcontrollers and compares security defenses on five real applications with 14 metrics, a different goal from finding weaknesses. IoTGoat \cite{owasp2026iotgoat} provides OpenWrt-based training firmware as binary images for virtual machines and selected physical boards, combining multiple weaknesses in one target. \gb combines binary firmware with reusable device profiles, simulated peripherals, and paired vulnerable and corrected variants of target weaknesses. Table~\ref{tab:comparison} compares them.

\begin{table}[ht]
\centering
\caption{IoT benchmarks compared with \gb.}
\label{tab:comparison}
\footnotesize
\adjustbox{width=0.9\columnwidth}{%
\begin{tabular*}{\columnwidth}{@{\extracolsep{\fill}}llll@{}}
\toprule
Benchmark & Software & Weaknesses & Size \\
\midrule
B4IoT \cite{10.1145/3600160.3600181} & own & features & container images \\ \rowsep
IoTBenchSL \cite{electronics14050856} & real & disclosed CVEs & 100 firmware \\ \rowsep
IoTFuzzBench \cite{electronics12143010} & real & 30 verified & 14 firmware \\ \rowsep
FirmBench \cite{duong2026firmrebugger} & real & 313 bug oracles & real binaries \\ \rowsep
BenchIoT \cite{8809492} & real & not applicable & 5 applications \\ \rowsep
IoTGoat \cite{owasp2026iotgoat} & OpenWrt & many, mixed & one platform \\ \rowsep
\gb & own & \makecell[l]{one primary CWE\\per vulnerable image} & \makecell[l]{\NImages vulnerable\\\NTwins corrected} \\
\bottomrule
\end{tabular*}%
}
\end{table}

\textit{Ground-truth benchmarks beyond IoT.} Juliet \cite{testjuliet} offers synthetic C/C++ and Java test cases with vulnerable and corrected variants.
LAVA \cite{7546498} injects bugs into real programs automatically, and Magma \cite{10.1145/3428334} forward-ports real bugs and scores fuzzers on reaching, triggering, and detecting them.
Ghera \cite{10.1145/3127005.3127010} is the closest in spirit, a small open repository of Android apps with one vulnerability each.
\gb applies the idea of one known target weakness per test case to running devices instead of source snippets or single programs.

\textit{Datasets for the MSR community.} AndroZoo \cite{allix2016androzoo} showed what a large shared corpus of real apps enables for mining studies, gaining scale and realism but giving up known ground truth.
A synthetic benchmark makes the opposite trade, a smaller, freely shareable artifact with known weaknesses.
\section{Data Availability}

The OpenEmbedded layer, the build and test scripts, and the documentation of every scenario are available at \url{https://github.com/Mobile-IoT-Security-Lab/IoTPwnBench}.
Researchers generate the binaries they need with \path{scripts/build-all.sh}, which builds the default vulnerable set on one machine.
With \texttt{--all --fixed}, it builds the full catalog, including opt-in families and corrected twins, on all five machines.
The build needs only Docker with Compose, and the repository lists the extra tools its host test scripts need.

Each build exports a kernel binary and a compressed ext4 filesystem image under \path{deploy-images/<firmware>/<machine>/}.
The documentation lists the boot command, payload, and signal of every scenario.

The code, recipes, and documentation are released under the MIT license in the top-level \texttt{LICENSE} file.
The two modules that bundle third-party code keep its license and declare it in their recipes.
The images are meant only for isolated laboratory use.
Upon acceptance we will archive a tagged release on Zenodo and cite its DOI in the final version.
\section{Conclusion}

We presented \gb, a synthetic benchmark of \NImages vulnerable firmware images and \NTwins corrected twins across \NDevices IoT device profiles.
Simulated peripherals and native services provide changing runtime behavior, and the binaries support static inspection and execution under QEMU.
Corrected twins retain the device and service without the target weakness, enabling paired evaluation of detection and false positives for that weakness.
Vulnerable images were validated manually, and the modular build regenerates them bit-for-bit.
Future work will evaluate analysis tools and add device classes.

%%%%% BIB %%%%%

\IEEEtriggeratref{14}
\bibliographystyle{IEEEtran}
\bibliography{references}

% END
\end{document}